\documentclass[aps,amsmath,amssymb,showpacs,preprint]{revtex4}

\usepackage{amsfonts}
\usepackage{amsmath}
\usepackage{amssymb}
\usepackage{graphicx}
\usepackage{color}

\newcommand{\avec}[1]{\mathbf{#1}}

\newcommand{\beq}{\begin{equation}}
\newcommand{\eeq}{\end{equation}}
\newcommand{\bsp}{\begin{split}}
\newcommand{\esp}{\end{split}}
\newcommand{\bpm}{\begin{pmatrix}}
\newcommand{\epm}{\end{pmatrix}}

\newcommand{\eq}[1]{Eq. [\ref{#1}]}
\newcommand{\fig}[1]{Fig.[\ref{#1}]}

\begin{document}

\title{Realising Haldane's vision for a Chern insulator in buckled lattices}

\author{Anthony R. Wright}
\email{a.wright7@uq.edu.au}
\affiliation{School of Mathematics and Physics, University of Queensland, Brisbane, 4072 Queensland, Australia}

\date{\today}

\begin{abstract}
The Chern insulator displays a quantum Hall effect with no net magnetic field. Proposed by Haldane over 20 years ago, it laid the foundation for the fields of topological order, unconventional quantum Hall effects, and topological insulators. Despite enormous impact over two decades, Haldane's original vision of a staggered magnetic field within a crystal lattice has been prohibitively difficult to realise. In fact, in the original paper Haldane stresses his idea is probably merely a toy model. I show that buckled lattices with only simple hopping terms, within in-plane magnetic fields, can realise these models, requiring no exotic interactions or experimental parameters. As a concrete example of this very broad, and remarkably simple principle, I consider silicene, a honeycomb lattice with out-of-plane sublattice anisotropy, in an in-plane magnetic field, and show that it is a Chern insulator, even at negligibly small magnetic fields, which is analogous to Haldane's original model.
\end{abstract}

\pacs{73.43.-f,73.43.Cd, 73.43.Jn}

\maketitle

\subsection*{Introduction}
F.D.M. Haldane's 1988 paper, `Model for a Quantum Hall Effect without Landau Levels: Condensed-Matter Realization of the ``Parity Anomaly''' \cite{hald}, was a paradigm shifting work which is undoubtedly a precursor to the far-reaching field of topological insulators and superconductors \cite{kane, zhangrev, kanerev}. In fact, topological insulators in an external field, or in the presence of ferromagnetic exchange coupling, are thought to be Chern, or quantum anomalous Hall (QAH) insulators \cite{QAH}, of which Haldane's model is the seminal example. 

An ultracold gas implementation of Haldane's model was recently proposed, which utilised laser induced pseudomagnetic fields \cite{new1} to produce a staggered flux \cite{ultracold2, gold}, and some progress has been made with these flux phases \cite{ultracold1}. The necessary Dirac physics has been suggested to be realizable in optical lattices \cite{delgado} due to the impressively tuneable gauge fields achievable with such systems. Realistic condensed matter realizations of staggered flux rely on interaction induced mean field flux phases, such as in the topological Mott insulator proposals \cite{TMI, TMI2, TMI3}. From a different route, the quantum anomalous Hall state was recently observed \cite{qah2013}. Since these foundational works, the notion of fractional Chern insulators \cite{frac, flat} has recently emerged. A partially filled, (almost) flat band in a Chern insulator can display fractional quantum Hall phenomena, i.e. fractionalized charge and statistics.

In the current work, I present a simple, general route to realising Chern insulator phases in lattice systems, relevant to condensed matter lattices, optical lattices, and indeed any periodic system. The only required ingredients to realise a Chern insulator, are the standard nearest and next-nearest neighbour hopping matrix elements, on a suitably buckled lattice. There is no requirement for interactions, order parameters, supercells, synthetic gauge fields, or even spin orbit coupling, making this, to the author's knowledge, the simplest and possibly the most practical approach to realising Chern insulators in real materials, and in fact does so via a \emph{real} staggered magnetic field.  

Specifically, I show that Haldane-like models can be constructed, by inducing a staggered flux that retains translational invariance, by employing an in-plane magnetic field incident on a buckled two dimensional system. Importantly, I show that such a system realizes the same chiral states as in the original model, and yet does not have the same chiral structure of flux accumulation.  The orientation of the incident field induces phase transitions between the Chern and ordinary insulating states. 

The current proposal is potentially quite broad, being applicable to many different systems in various contexts. However, there are two essential ingredients in any specific application. Firstly, a buckled lattice is required, such that a staggered magnetic field pattern can be obtained, while no net magnetic field passes through the unit cell. Secondly, the dispersion relation in the absence of an in-plane magnetic field, must contain symmetry-protected gapless Dirac points, or at least, those with band-gaps which are smaller than those induced by the magnetic flux.

\begin{figure}[t]
\centering\includegraphics[width=7cm]{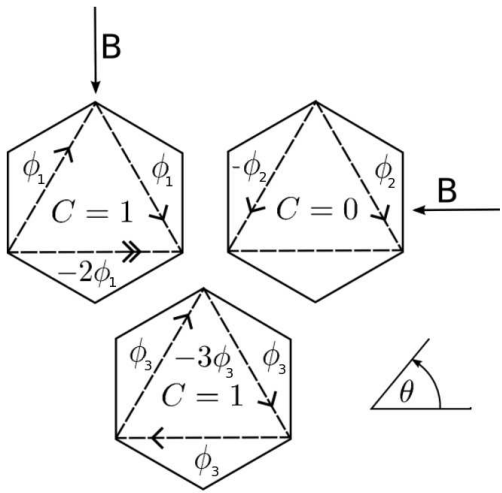}
\caption{Hexagonal unit cells with zero net magnetic flux. The upper two unit cells are a buckled honeycomb lattice in an in-plane magnetic field. In the lower pane is shown Haldane's unit cell \cite{hald}. The direction of positive flux accumulation is indicated by the arrows along the bonds. For clarity, I have only shown the bonds along a single sublattice.}
\label{flux}
\end{figure}

\subsection*{Results}
\subsubsection*{Prototypical model}
I choose silicene as the prototype buckled lattice \cite{sili0}. Silicene is a two dimensional honeycomb lattice, where the two sites in the unit cell are offset from each other, in the out-of-plane direction. I define the plane parallel to the lattice, $z = 0$, such that one triangular sublattice is at $z = l$, and the other at $z = -l$. Ezawa has introduced an unusual quantum anomalous Hall effect for silicene under circularly polarized light. The breaking of time reversal symmetry in this case is achieved by the time dependence of the electric field, and the topological phase is a nonequilibrium one, arising from Floquet theory \cite{EZPRL2013}

The silicene Hamiltonian in an in-plane magnetic field and an out-of-plane electric field then, is \cite{sili}:

\beq
\bsp
H &= t\sum_{\langle i,j\rangle,}c_{i}^\dag c_{j} +\sum_{\langle\langle i,j\rangle\rangle,}t'_{ij}c_{i}^\dag c_{j} + \Delta\sum_{i,} \nu(i)c_{i}^\dag c_{i},
\end{split}
\label{H}
\eeq
where $t$ is the nearest neighbour hopping integral, $t'_{ij} = t'\exp(i\frac{e}{\hbar}\int_i^j \vec{A}\cdot d\vec{l})$ is the Peierls' substitution modified next nearest neighbour hopping, $\Delta = lE_z$ is the electric field multiplied by the out-of-plane buckling $l = 0.23\AA$. The out-of-plane buckling is \emph{essential} to the current proposal, as it allows a non-trivial flux pattern, as shown in \fig{flux}, to penetrate the unit cell. For this reason, a flat graphene sheet is not suitable to realise the proposed effect. The out-of-plane field is not necessary to observe the QAH effect, but allows us to make contact with Haldane's work \cite{hald}, and when including spin effects, becomes important. I have chosen a gauge such that the flux is accounted for on next nearest neighbour bonds only. With the two sublattices at $z = \pm l$, this is the Landau gauge, $\vec{A} = Bz\bigl(\sin\theta, -\cos\theta, 0)$, where $\theta$ is the angle of the field relative to the horizontal direction in \fig{flux}.

In momentum space, the Hamiltonian becomes

\beq
H_{\vec{k}} = \vec{c}_{\vec{k}}^\dag \bigl(d_0(\vec{k})\sigma_0 + \vec{d}(\vec{k})\cdot \vec{\sigma}\bigr)\vec{c}_{\vec{k}},
\label{HK}
\eeq
where $\vec{c}_{\vec{k}}^\dag = (c_{A,\vec{k}}^\dag,c_{B,\vec{k}}^\dag)$, with $A,B$ denoting the honeycomb sublattices, and in which 

\beq
d_0(\vec{k}) = \sum^3_i 2t' \cos(\frac{\phi}{\phi_0}\vec{a}_\theta\cdot\delta'_i)\cos(\vec{k}\cdot\hat\delta'_i),
\eeq
where $\phi_0 = h/2e$, and $\phi/\phi_0$ is the integrated gauge field over the bond using the Landau gauge, which gives the total flux through the loop with one next nearest neighbor and two nearest neighbor bonds, as indicated in \fig{flux}, and $\vec{a}_\theta = (\sin\theta, -\cos\theta)$ is the two-dimensional unit vector parallel to our chosen vector potential $\vec{A}$. $\delta'_i$ are three out of six of the next nearest neighbor vectors defined by $\delta'_n = R(2\pi n/3)(1,0)$, where $R(\theta)$ rotates the vector by $\theta$, and

\beq
\bsp
\vec{d}_i(\vec{k}) = \bigl(&t \cos(\vec{k}\cdot\delta_i), t\sin(\vec{k}\cdot\delta_i), \Delta-2t'\sin(\frac{\phi}{\phi_0}\vec{a}_\theta\cdot\delta'_i)\sin(\vec{k}\cdot\hat\delta'_i)\bigr),
\end{split}
\label{dveceq}
\eeq
where $\delta_i$ are the three nearest neighbor vectors defined by $\delta_n = R(2\pi n/3)(0,-1/\sqrt{3})$. Here and throughout, we take the next nearest neighbour vector to determine the natural length scale and set it to $1$ such that all momenta are dimensionless. 

The key difference between our buckled lattice Hamiltonian and that proposed by Haldane \cite{hald}, is the vector $\vec{a}$, which modulates the next nearest neighbour flux accumulation. In the original work, a magnetic flux configuration was chosen such that the flux accumulation along \emph{all} next nearest neighbour hops was equal. Here it is strongly angle dependent, and differs in sign for different hops. Nevertheless, an equivalent chirality emerges due to the breaking of time reversal symmetry, as we shall see. 

The full Hamiltonian, re-introducing the dependence on all variables, is given by

\beq
H(\vec{k})_{\theta,\phi} = 
\bpm
\Delta + h_D(\vec{k})_{\theta,\phi} & e^{-i\frac{k_y}{\sqrt{3}}} + 2\cos(\frac{k_x}{2})e^{i\frac{k_y}{2\sqrt{3}}}\\
e^{i\frac{k_y}{\sqrt{3}}} + 2\cos(\frac{k_x}{2})e^{-i\frac{k_y}{2\sqrt{3}}} & -\Delta + h_D(\vec{k})_{\theta,\phi}
\epm,
\label{hfull}
\eeq 
where $h_D(\vec{k})_{\theta,\phi} = 2t'\bigl(2\cos(\frac{\sqrt{3}}{2}(k_y - \phi\cos\theta))\cos(\frac{1}{2}(k_x + \phi \sin\theta)) + \cos(k_x + \phi\sin\theta)\bigr)$.

\begin{figure}[t]
\centering\includegraphics[width=8cm]{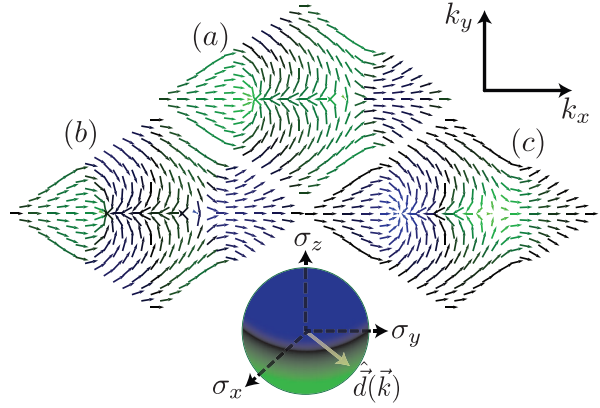}
\caption{The unit vector $\vec{d}(\vec{k})$, of \eq{dveceq}, plotted across the honeycomb lattice Brillouin zone. Due to the periodic boundary conditions of the Brillouin zone, the unit vector must wrap the sphere an integer number of times, which gives the Chern number. In all three figures, $t' = 0.1t$. In (a), $\Delta = 0.2t$, and so the system is a trivial insulator. The unit vector clearly never visits the north pole, but wraps and then un-wraps the lower hemisphere. In (b), $\Delta = 0$, $\phi = \pi/2\phi_0$, and $\theta = \pi/2$, so the system is a Chern insulator. In (c) is shown the Haldane model. In the later two cases, it can be seen that the vector visits both the north and south poles once only, giving a Chern number of 1. The bottom figure shows the representation of the unit vector in pseudo-spin space, together with the corresponding colour coding of the $\sigma_z$ component of each vector.}
\label{dvec}
\end{figure}

\subsubsection*{Flux gaps and chirality}
The system, together with the inclusion of flux along the bonds, and the comparison with Haldane's model, are all shown in \fig{flux}. The chirality, noted by Haldane in his model, is the clockwise accumulation of positive flux over next nearest neighbor hops. The clockwise accumulation of positive flux is clear in the lower panel in \fig{flux}. This chirality is absent in the buckled lattices (i.e. the upper two unit cells shown in \fig{flux}), where hopping over the equivalent loop leads to zero phase accumulation. However, due to the particular out-of-plane buckling of the silicene lattice, a loop comprising one next nearest neighbor and two nearest neighbor hops acquires a non-zero positive phase. For the upper left panel, there are two loops which acquire a negative phase on clockwise hopping, and one which acquires a positive phase. There is, therefore, a net chirality. In the upper right however, the lower triangular plaquette envelopes no magnetic flux, and thus there are just two oppositely oriented chiral loops, and thus no net chirality. I mention that very similar flux configurations, arising from proposed laser induced pseudomagnetic fluxes in ultracold atom systems, have recently been discussed, complementing the current study \cite{gold}. 

With all fields off, the honeycomb lattice has two low energy massless Dirac cones at the two inequivalent $K$ points, $K_\pm = (\pm 4\pi/3, 0)$. Near these two inequivalent valleys, we can expand for small $\vec{k}$, and obtain the low energy theory of our buckled lattice, given by

\beq
H_\pm(\vec{k})_{\theta,\phi} \approx -3t'\sigma_0 + v_F k_x\sigma_x + v_F k_y \sigma_y + \bigl(\frac{3}{2}\frac{\phi}{\phi_0} t'(k_y\cos\theta - k_x\sin\theta) \pm \Delta_{\theta,\phi}\bigr)\sigma_z 
\eeq 
for small $\phi$, where $v_F$ is the usual Fermi velocity for a honeycomb lattice, $v_F = \sqrt{3}t/2$.

$\Delta$ and $\vec{B}$ then introduce separate mass terms. The $\Delta$ mass term is due to the on-site energy imbalance between the two sublattices due to the electric field Stark effect. The magnitude of this gap is $2El$. The second, which I call the `flux gap', is the sublattice anisotropy due to the orbital effect of the magnetic field, and its magnitude is 

\beq
\Delta_{\theta,\phi}=\bigl|4\sqrt{3}t'\sin(\frac{\phi}{2}\sin(\theta))\bigl(\cos(\frac{\phi}{2}\sin(\theta)) - \cos(\frac{\sqrt{3}\phi}{2}\cos(\theta))\bigr)\bigr| . 
\label{deltac}
\eeq
For small $\phi$, this gap goes as $\phi^3$, being given by

\beq
\Delta_{\theta,\phi} \approx \frac{\sqrt{3}}{4}t'\sin(3\theta)\biggl(\frac{\phi}{\phi_0}\biggr)^3,
\eeq
and so will be tiny for realistic laboratory fields. Later, I will discuss ways in which this number can be improved by several orders of magnitude in the lab. $\Delta$ and $\Delta_{\theta,\phi}$ add, or compete, in the two valleys such that the gap in the two valleys, are $\Delta_{K_\pm} = \Delta \pm \Delta_{\theta,\phi}$.

\subsubsection*{Chern number and phase diagrams}
The quantum anomalous Hall (QAH) phase is characterised by a Chern number of 1 ($\mathrm{Mod}(2)$), where the Chern number is the integral of the Berry curvature over the Brillouin zone \cite{chern}. For our Hamiltonian \eq{H}, it takes the particularly simple form \cite{zhangrev, moore2}

\beq
\bsp
C &= \frac{1}{4\pi}\int_{BZ} d\avec{k}\hat{d}\cdot\frac{\partial \hat{d}}{\partial k_x}\times\frac{\partial \hat{d}}{\partial k_y} = \left\{
\begin{array}{rl}
1 &\,\,\,|\Delta_{\theta,\phi}|>|\Delta|\\
0 & \,\,\,|\Delta_{\theta,\phi}| < |\Delta|,
\end{array} 
\right.
\end{split}
\label{chern}
\end{equation}
where $\hat{\vec{d}}(\vec{k}) = \vec{d}(\vec{k})/|\vec{d}(\vec{k})|$.

\begin{figure}[t]
\centering\includegraphics[width=8cm]{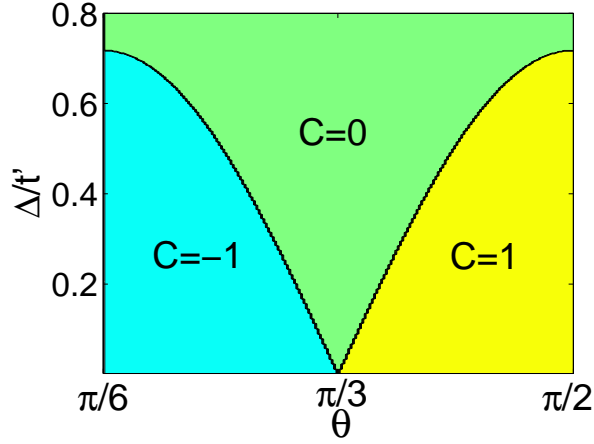}
\caption{Phase diagram for a spinless buckled honeycomb lattice in an in-plane magnetic field (such that $\phi/\phi_0 = \pi/2$), and out-of-plane electric field, where $\Delta = El$. The angle $\pi/2\,(\pi/3)$ corresponds to the upper left (right) geometry in \fig{flux}.}
\label{phasediagram}
\end{figure}

The Chern number discriminates between ostensibly similar, or even identical, bandstructures, by calculating the integral of the Berry curvature \cite{berry} over the Brillouin zone. For two component Hamiltonians, such as \eq{hfull}, the Chern number calculation is readily visualised across the Brillouin zone. By considering the orientation of the unit vector $\hat{\vec{d}}(\vec{k})$ at each point, and specifically, noting how many times the vector wraps around the unit sphere, one can immediately `read off' the Chern number. In \fig{dvec}, I have plotted the unit vector $\hat{\vec{d}}(\vec{k})$ over the hexagonal lattice Brillouin zone for the original Haldane model with no mass gap and finite flux ($C=1$), as well as \eq{hfull} with $\Delta = 0.2t$, $\phi = 0$, which is a trivial insulator ($C=0$), and $\Delta = 0$, but $\phi/\phi_0 = \pi/2$, $\theta = \pi/2$, which is again a Chern insulator ($C=1$), being in the same topological class as the Haldane model. Remarkably, the bulk bandstructure in all three cases are \emph{nearly} identical, being simply essentially graphene \cite{gappedgraphene}, and so cannot be distinguished by viewing the bandstructure alone.

In \fig{phasediagram} is shown the Chern number phase diagram of Hamiltonian \eq{H}, as a function of in-plane field orientation $\theta$ (as defined in \fig{flux}). At $\Delta = 0$, the Chern number is $\pm1$ for \emph{almost} the entire spectrum of field orientations. This can be understood by considering \fig{flux}. For all $\theta$ except $\pi n/3$, there is a positive flux passing through either one or two plaquettes in the unit cell, and a negative flux passing through two or one, such that the total flux is always zero. The positive fluxes induce a positive chirality about the triangular plaquettes in one direction, while the negative flux induce a negative chirality. Therefore we expect there to be a net chirality inducing a chiral mode at each edge of the material.  A topological phase transition can be induced by varying the orientation of the in-plane magnetic field. It is also worth noting that at $\Delta = 0$, with finite $t'<t/3$, and at zero temperature, the system is a quantum anomalous Hall insulator for \emph{any} nonzero magnetic field magnitude, so long as the field orientation is not precisely $\theta = \pi n/3$.

As can be seen from \fig{flux}, the field orientation angles $\theta = \pi n/3$ are special in that a single triangular plaquette lies in a plane parallel to the field and thus sees no net flux through it. The sine dependence of the critical gap can be understood by noting that the flux passing through a plaquette goes as the sine of the angle between the plaquette and the field. For a decreased flux through a plaquette, the critical electric field to destroy the chiral edge modes is also decreased.

Although clearly distinct from the Haldane model in its flux configuration, the buckled honeycomb lattice in an in-plane magnetic field is \emph{topologically equivalent} to the Haldane model for ranges of  magnetic field orientations, as both systems have a Chern number of one.

\subsubsection*{Effects of spin-orbit coupling}
So far I have completely neglected spin. The Chern numbers reported in the phase diagram \fig{phasediagram} for a fermionic system are \emph{per spin}. For spin degenerate systems such as that considered here, each spin species will co-propagate. The Zeeman splitting will not affect the Chern number, but if it is larger than the flux gap it will move the bulk bands of one spin species across the Fermi energy, thus developing a Fermi surface. Experimentally, this is an added complication, and will be addressed briefly toward the end of the paper. Therefore, rather than having a Chern number in Hamiltonian \eq{H} of 1, for a spinful fermionic system, there is an extra factor of 2 for the spin degeneracy, and the Chern number is in fact 2, corresponding to a $\nu = 2$ quantum Hall effect, with 2 filled Landau levels. I emphasise that this is not a quantum spin Hall insulator, where the total Chern number is $C = 1-1 = 0$ and the two spin species counter-propagate, but is a doubled Haldane model, or a quantum Hall effect with filling factor 2. Back scattering on the edges is still prohibited in this model, as there is only one direction of propagation on the edge.

I now add a spin index to \eq{H}, and introduce the Kane-Mele spin orbit coupling term between next nearest neighbors \cite{kane}, 

\begin{figure}[t]
\centering\includegraphics[width=8cm]{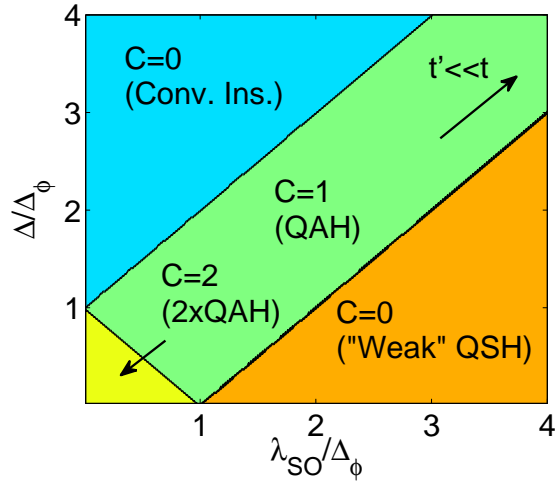}
\caption{Spinful phase diagram of silicene. For small spin orbit coupling and small electric field, we obtain two co-propagating chiral bands, or a $\nu=2$ QHE. For $\Delta$ larger than the flux gap, the system iss a trivial insulator. For comparatively large spin orbit coupling, silicene is a Kane-Mele topological insulator \cite{kane}. When $\lambda_{SO}+\Delta>\Delta_{\theta,\phi}$, and $|\lambda_{SO} - \Delta|<\Delta_{\theta,\phi}$, silicene is a Chern insulator. The diagonal $C = 1$ section continues indefinitely, so long as the energy scales $\Delta\sim\lambda_{SO}\ll t$. Note that in the weak-QSH phase there are helical edge states, yet there is also a magnetic field that weakens the robustness against backscattering, and thus the classification of `weakÕ-QSH.}
\label{phasediagram2}
\end{figure}

\beq
H_{SO} = \frac{\lambda_{SO}}{3\sqrt{3}}\sum_{\langle\langle i,j\rangle\rangle,\sigma,\sigma'}\nu_{ij}c_{i\sigma}^\dag\sigma_{\sigma\sigma'}^z c_{j,\sigma'},
\eeq
where $\nu_{ij} = \pm1$ in an alternating fashion. In silicene, $\lambda_{SO} \approx 4$ meV\cite{ezawaphoto}. The induced mass has different signs in both the valley and spin degrees of freedom. Therefore it modifies the gap $\Delta_{K_\pm}$ such that \cite{ezawa16} $\Delta_{K_\pm,s} = \Delta \pm\Delta_{\theta,\phi} \pm s\lambda_{SO}$, where $s = \pm$ corresponds to spin. A lucid analysis, overtly exploring the link between the signs of mass terms with the edge state spectrum, was recently conducted \cite{gapsign1, gapsign2}.

I mention briefly at this point that the Rashba spin orbit interaction contributes also to the results that follow, especially in the presence of an inversion symmetry breaking field. However, the effect is relatively small compared with the intrinsic spin orbit interaction, and does not, therefore, for Rashba interaction strengths expected in silicene \cite{ezawa16}, significantly alter the phase diagram. Nevertheless, for a system with small flux gaps, and zero intrinsic spin orbit interaction, the Rashba interaction can destroy the chirality in an in-plane magnetic field, and Rashba terms must be included in this case.

When $\lambda_{SO}+\Delta>\Delta_{\theta,\phi}$, a mass inversion occurs \emph{for a single spin species, in a single valley}. The Chern number becomes 1. From the phase diagram \fig{phasediagram}, the way to understand this is that one species has mass $\Delta_{\theta,\phi} > \Delta - \lambda_{SO}$, and thus the contribution to the Chern number from that band is $1$, whereas the other has $\Delta_{\theta,\phi} < \Delta + \lambda_{SO}$, and thus the contribution from that band is 0. For larger spin orbit couplings however, a second mass inversion takes place, when $|\Delta - \lambda_{SO}|>\Delta_{\theta,\phi}$, such that the system becomes equivalent to the original quantum spin Hall insulator \cite{kane}, with counter-propagating edge states, and a Chern number of zero. These phases are all indicated in the spinful phase diagram, \fig{phasediagram2}.

\subsubsection*{Edge states and the bulk-edge correspondence}
Having obtained the phase diagrams for both the spin-less and spin-ful cases by calculating the corresponding Chern numbers of each \emph{bulk} band, we can immediately predict the \emph{edge} state structure of our system in each phase. This can be achieved by the so-called `bulk-edge correspondence' \cite{edge, edge2}. Referring to \fig{dvec}, we note the following observations. Firstly, the wrapping of the unit vector $\hat{d}(\vec{k})$ around the unit sphere, determines the Chern numbers of the two bands. Specifically, over the full Brillouin zone, if the unit vector wraps around the unit sphere an integer $n$ number of times, then the conduction band has Chern number $C_{c}=\pm n$, and the valence band has Chern number $C_{v} = \mp n$, where the $\pm$ is determined by the sense of the wrapping, which is not relevant to us here. By continuity of $\hat{d}(\vec{k})$ together with the periodic boundary conditions of the Brillouin zone up to a unitary phase (i.e., the Brillouin zone is topologically equivalent to a 2-torus), we are assured that the wrapping of the unit vector $\hat{d}(\vec{k})$ can only take integer values. Secondly, we note that there is no continuous way to deform the function $\hat{d}(\vec{k})$ over the Brillouin zone, such that the wrapping of the unit sphere changes by an integer. This point leads directly to the bulk-edge correspondence. Namely, if a system with Chern number one is placed next to a system with Chern number two, then the winding of the unit vector around the unit sphere over the Brillouin zone must change \emph{abruptly} at some point near the interface of the two systems. This can only happen if the unit vector $\hat{d}(\vec{k})$ \emph{vanishes}. Referring to \eq{HK}, we see that the Hamiltonian becomes diagonal at that point, and therefore corresponds to a \emph{degeneracy}. These points are discussed in much more rigorous detail elsewhere \cite{edge, edge2}.

Armed with the insight that the unwrapping of the unit vector around the unit sphere corresponds to a degeneracy point, we are assured then that near the edge of a finite slab of our honeycomb ribbons, if the Chern number is one, then there must be a state near each edge which crosses the chemical potential.  

\begin{figure}[t]
\centering\includegraphics[width=10cm]{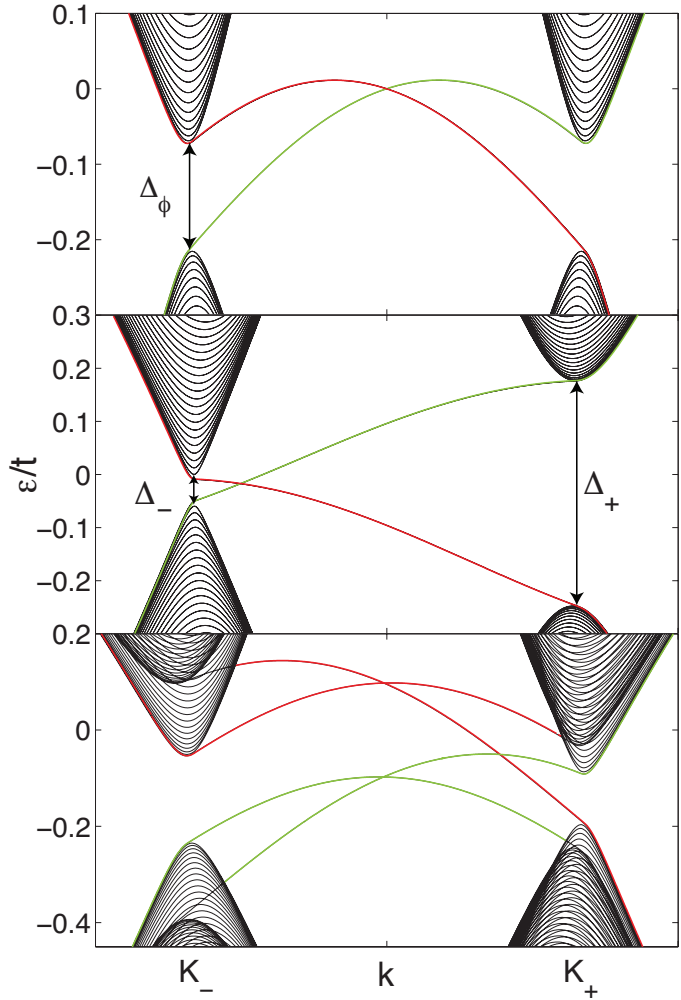}
\caption{Dispersion of energy eigenstates for ribbons of: spinless silicene (upper) with $\Delta = 0$, the Haldane model (middle) with $\Delta = 0.05t$, and spinful silicene with $\lambda_{SO} = 0.07t$, $\Delta = 0.05t$ (lower). Both the Haldane model and spinless silicene have a single chiral mode \emph{per ribbon edge}. Therefore the Chern number in the two systems is the same. In the lower pane, $\lambda_{SO}$ has caused a band inversion for one K-point, and one pseudo-spin, such that one chiral mode is destroyed, but the other survives, and the Chern number is 1. The other parameters for the silicene systems: $t' = 0.1t$, $\phi/\phi_0 = \pi/2$, and $\theta = \pi/2$. In each figure, the edge states are coloured, with red depicting states bound to one edge of the ribbon, and green depicting the opposite edge.} 
\label{ribfig}
\end{figure}

In \fig{ribfig}, the energy band structure obtained by solving the model on a ribbon \cite{edge} is shown for three different configurations. For spinless silicene, there is a
single chiral edge mode crossing the bulk gap. Edge modes corresponding to opposite edges of the ribbon have opposite chirality. The Haldane model is also shown in the middle pane, which just has a single propagation direction per edge band. The lower panel shows the effect of the spin orbit coupling on the edge states. The spin orbit interaction inverts the mass of a single spin species at a single valley, leading to the destruction of a single chiral edge state, but not the other, and thus the Chern number is 1. Note that the gap at the $K_\pm$ points are given by the sum and difference of the constituent gaps.

\subsubsection*{Experimental considerations}
Finally, I estimate the magnetic field needed to produce a Haldane-like model in a real system. For any finite magnetic field, at zero temperature, silicene is a Chern insulator. In silicene, the next nearest neighbor bonds are separated by $\approx 4\,\AA$, and the vertical buckling amount is $\approx 0.46\,\AA$. Therefore $\phi/\phi_0\approx 2\times10^{-5}B\,\mathrm{T}^{-1}$. Unfortunately, for small fields, the flux-gap \eq{deltac} goes as $\phi^3$, as mentioned earlier, and so an extremely large field ($\sim 100$ T) is required to obtain a gap  of only $2\mu K$. 

A gap which goes as $\phi$ can be achieved with next nearest neighbor anisotropy, for example, by the inclusion of a plane polarized laser, which modifies the hopping by a time averaged gauge field. In the regime $\hbar\omega\ll \Delta_{K_-}$, $t\rightarrow J_0(eaE/\hbar\omega)t$ \cite{bessel,inoue}, where $\vec{E} = E_0\cos(\omega t)\hat{E}$, and $a$ is the lattice constant. This modifies the gap equation \eq{deltac} such that, for an electric field polarized in the plane parallel to $\theta = 0$, and for small $\phi$ 

\beq
\Delta_{\theta,\phi} = \sqrt{3}t'\phi/\phi_0|(J_0(eaE_0/2\hbar\omega) - J_0(eaE_0/\hbar\omega))\sin(3\theta)|,
\label{deltacp}
\eeq
where $J_0$ is the modified Bessel function of the second kind. As a simple example, with laser frequency $\hbar\omega = 10$meV, at an intensity of $5\times10^7$ Vm$^{-1}$, a flux gap of $\Delta_{\theta,\phi} \approx 1.3 t'\phi/\phi_0$ is achievable, which, for a $35$ T field, is $\approx 0.5$ K. 

I stress, however, that this is not a necessary application of Floquet theory. The key requirement is the breaking of the C3 rotational symmetry of the hexagonal lattice. It is this rotational symmetry that forces the flux gap to be prohibitively small. While a plane polarized electric field may be the simplest way to break this symmetry, it is by no means the only way. As an alternative example, under a simple uni-axial strain, the flux gap is modified as $\Delta_{\theta,\phi} \propto \delta t' \phi$, where $t'\rightarrow t'+\delta t'$ along the strain axis. However, possible emergent gauge fields must be considered in this case \cite{gauge}.

In the above, the Zeeman splitting will likely cause bulk bands for one K point to cross the Fermi energy, thus developing a Fermi surface. For a g-factor of 2, the Zeeman splitting at 35 T is $\sim 20$ K, which in the above, exceeds the flux gap. The Zeeman splitting does not affect the Chern number. Away from the K points, the bulk bands will still be gapped in general, and so the gapless edge states could feasibly be imaged with angle resolved photoemission spectra. Alternatively, the edge states could be measured via a nonlocal transport experiment, using an H-shaped bar type geometry, precisely the way in which 2D topological insulator edge states have been compellingly verified \cite{roth}.

To maximize the staggered flux through the unit cell, bilayer systems can be envisaged where the inter-layer gap is much larger than that reported here, due to the vastly increased interlayer spacing. Alternatively, ``designer Dirac fermions'' engineered by precise placement of carbon monoxide molecules to produce ``molecular graphene'' \cite{designer} could be engineered to produce a buckled structure with large out-of-plane displacement, and/or strong next nearest neighbor coupling and anisotropy. 

\subsection*{Methods}
Crystal structures were calculated using standard tight binding techniques. The Peierl's substitution in the tight-binding context was adopted in order to incorporate the external magnetic field. As the field is in-plane, orbital effects are irrelevant and so Landau levels do not form. The gauge choice, centred about the vertical mid-point of the lattice is a Landau-type one, which was chosen for calculational convenience, and to highlight the role of the out-of-plane buckling. Chern number calculations were performed by numerical integration, using \eq{chern}. When including Rashba spin orbit interaction, the $SU(2)\times SU(2)$ structure breaks down to $SU(4)$, and so the standard numerical form of the Berry curvature was integrated over the Brillouin zone instead \cite{berry}. Ribbon dispersions were calculated within the same tight-binding formalism, by expanding the unit cell to cover the entire ribbon width, and keeping translational invariance along only the ribbon axis. The ribbons are all of zig-zag edge type.


\subsection*{Discussion}
I have introduced a new conceptual framework whereby Haldane-like models can be realized. The key material property in this proposal is a buckled lattice, such that an in-plane magnetic field produces a unit cell with zero net flux, yet regions with positive and negative flux which individually support cyclotron orbits. In this way, the chirality necessary in implementing a Haldane-like model can be achieved. Using silicene as a specific example, I showed that over a range of field angles, silicene in an in-plane field is topologically equivalent to the Haldane model per spin, with chiral edge states arising by similar, yet distinct physical processes, and that topological phase transitions can be induced by rotating the orientation of the field. Upon successful observation of a Chern insulator, engineering a flat band in the Brillouin zone could be investigated to explore the possibility of a fractional Chern insulator \cite{flat}.

\acknowledgments
I thank Ross McKenzie and Jure Kokalj for critical readings of the manuscript and helpful discussions. I also thank Jacopo Sabbatini, Omri Bahat-Treidel, Janani Chander and Michael Holt for enlightening discussions. I am financially supported by a University of Queensland Postdoctoral Fellowship.


\begin{thebibliography}{99}

\bibitem{hald}Haldane, F.D.M. Model for a Quantum Hall Effect without Landau Levels: Condensed-Matter Realization of the ``Parity Anomaly'', \emph{Phys. Rev. Lett.} \textbf{61,} 2015-2018 (1988).

\bibitem{kane}  Kane, C.L. and  Mele, E.J. Quantum Spin Hall Effect in Graphene,  \emph{Phys. Rev. Lett.} \textbf{95,} 226801-226803 (2005).

\bibitem{zhangrev}  Qi, X.-L. and Zhang, S.-C. Topological insulators and superconductors, \emph{Rev. Mod. Phys.} \textbf{83,} 1057-1110  (2011).

\bibitem{kanerev}  Hasan, M. Z. and  Kane, C. L. Colloquium: Topological insulators, \emph{Rev. Mod. Phys.} \textbf{82,} 3045Ð3067 (2010).

\bibitem{QAH}  Qi, X.-L.,   Wu, Y. S. and  Zhang, S. C. Topological quantization of the spin Hall effect in two-dimensional paramagnetic semiconductors, \emph{Phys. Rev. B} \textbf{74,} 085308-085314 (2006)

\bibitem{new1} Jaksch, D. and  Zoller, P. Creation of effective magnetic fields in optical lattices: the Hofstadter butterfly for cold neutral atoms, \emph{New J. Phys.} \textbf{5,} 56.1-56.11 (2003)

\bibitem{gold}  Goldman, N., et al. Measuring topology in a laser-coupled honeycomb lattice: from Chern insulators to topological semi-metals,  \emph{New J. Phys.} \textbf{15,} 013025-013054 (2013).

\bibitem{ultracold2}  Alba, E., et al. Seeing Topological Order in Time-of-Flight Measurements, \emph{Phys. Rev. Lett.} \textbf{107,} 235301-235305 (2011).

\bibitem{ultracold1}  Aidelsburger, M., et al. Experimental Realization of Strong Effective Magnetic Fields in an Optical Lattice, \emph{Phys. Rev. Lett.} \textbf{107,} 255301-255305 (2011).

\bibitem{delgado} Goldman, N., et al. Non-Abelian Optical Lattices: Anomalous Quantum Hall Effect and Dirac Fermions, \emph{Phys. Rev. Lett.} \textbf{103,} 035301-035304 (2009).

\bibitem{TMI} Raghu, S.,  Qi, X.-L.,  Honerkamp, C. and Zhang, S.-C., Topological Mott Insulators, \emph{Phys. Rev. Lett.} \textbf{100,} 156401-156405 (2008).

\bibitem{TMI2}  Dauphin, A.,  M\"uller, M.  Martin-Delgado, M. A. Rydberg-atom quantum simulation and Chern-number characterization of a topological Mott insulator, \emph{Phys. Rev. A} \textbf{86} 053618-053634 (2012).

\bibitem{TMI3}  Sun, K.,  Liu W. V. , Hemmerich, A. and Das Sarma, S. Topological semimetal in a fermionic optical lattice, \emph{Nat. Phys.} \textbf{8,} 67-70 (2011).

\bibitem{qah2013} Chang, C.-Z. et al. Experimental Observation of the Quantum Anomalous Hall Effect in a Magnetic Topological Insulator, \emph{Science}, \textbf{340,} 167-170 (2013).

\bibitem{frac}  Regnault, N. and  Bernevig B. A. Fractional Chern Insulator, \emph{Phys. Rev. X} \textbf{1,} 021014-021027 (2011).

\bibitem{flat}  Neupert, T.,  Santos,  L.,   Chamon, C. and  Mudry, C. Fractional Quantum Hall States at Zero Magnetic Field, \emph{Phys. Rev.  Lett.} \textbf{106,} 236804-236807 (2011).

\bibitem{BHZ}  Bernevig, B. A.,  Hughes, T. L. and  Zhang, S.-C. Quantum Spin Hall Effect and Topological Phase Transition in HgTe Quantum Wells, \emph{Science} \textbf{314,} 1757-1761 (2006)

\bibitem{sili0} Takeda, K., and Shiraishi, K. Theoretical possibility of stage corrugation in Si and Ge analogs of graphite, \emph{Phys. Rev. B} \textbf{50,} 14916Ð14922 (1994).

\bibitem{EZPRL2013} Ezawa, M. Photoinduced Topological Phase Transition and a Single Dirac-Cone State in Silicene, \emph{Phys. Rev. Lett.} \textbf{110,} 026603-026607 (2013).

\bibitem{sili} Liu, C.-C., Feng,  W. and  Yao, Y. Quantum Spin Hall Effect in Silicene and Two-Dimensional Germanium, \emph{Phys. Rev. Lett.} \textbf{107,} 076802-076805 (2011).

\bibitem{chern} Xiao,D.,  Chang, M.-C. and  Niu, Q. Berry phase effects on electronic properties, \emph{Rev. Mod. Phys.} \textbf{82,} 1959Ð2007 (2010)

\bibitem{ezawaphoto}  Ezawa, M. Spin-valley optical selection rule and strong circular dichroism in silicene, \emph{Phys. Rev. B} \textbf{86,} 161407-161410(R) (2012).

\bibitem{moore2} Cho, G.Y. and  Moore, J.E. Quantum phase transition and fractional excitations in a topological insulator thin film with Zeeman and excitonic masses, \emph{Phys. Rev. B} \textbf{84,} 165101-165110 (2011)


\bibitem{berry} M.V. Berry, Quantal Phase Factors Accompanying Adiabatic Changes, Proc. R. Soc. Lond. A \textbf{392,} 45-57 (1984).

\bibitem{gappedgraphene} G.W. Semenoff, Condensed-Matter Simulation of a Three-Dimensional Anomaly, \emph{Phys. Rev. Lett.} \textbf{53,} 2449-2452 (1984).


\bibitem{ezawa16} Ezawa, M. Valley-Polarized Metals and Quantum Anomalous Hall Effect in Silicene, \emph{Phys. Rev. Lett.} \textbf{109,} 055502-055506 (2012).

\bibitem{gapsign1} Goldman, N., Beugeling W. and Smith, C. M., Topological phase transitions between chiral and helical spin textures in a lattice with spin-orbit coupling and a magnetic field, \emph{Europhys. Lett.} \textbf{97,} 23003 (2012).

\bibitem{gapsign2} Beugeling, W.,  Goldman, N. and  Smith, C. M., Topological phases in a two-dimensional lattice: Magnetic field versus spin-orbit coupling, \emph{Phys. Rev. B} \textbf{86,} 075118-075135 (2012).

\bibitem{edge} Hatsugai, Y. Chern number and edge states in the integer quantum Hall effect, \emph{Phys. Rev. Lett.} \textbf{71,} 3697-3700 (1993). 

\bibitem{edge2} Hatsugai, Y. Edge states in the integer quantum Hall effect and the Riemann surface of the Bloch function, \emph{Phys. Rev. B} \textbf{48,} 11851Ð11862 (1993).

\bibitem{bessel} Eckardt, A., Weiss, C. and Holthaus, M. Superfluid-Insulator Transition in a Periodically Driven Optical Lattice, \emph{Phys. Rev. Lett.} \textbf{95,} 260404-260407 (2005).

\bibitem{inoue} Inoue, J.I. and Tanaka,  A. Photoinduced Transition between Conventional and Topological Insulators in Two-Dimensional Electronic Systems, \emph{Phys. Rev. Lett.} \textbf{105,} 017401-017404 (2010).

\bibitem{gauge} Levy, N., et al. Strain-Induced PseudoÐMagnetic Fields Greater Than 300 Tesla in Graphene Nanobubbles, \emph{Science} \textbf{329,} 544-547 (2010).

\bibitem{designer} Gomes,  K. K.,  Mar, W., Ko, W., Guinea,  F. and Manoharan, H.C. Designer Dirac fermions and topological phases in molecular graphene,  \emph{Nature} \textbf{483,} 306-310 (2012).

\bibitem{roth} Roth, A., et al.  Nonlocal Transport in the Quantum Spin Hall State, \emph{Science} \textbf{325,} 294-297 (2009).













\end{thebibliography}
\end{document}